\documentclass[showpacs,onecolumn,floats,superscriptaddress]{revtex4}
\usepackage{amssymb}

\usepackage{graphicx}

\begin{document}

\title{Collective Modes in Two-band Superconductors.\\
}
\author{A. Anishchanka$^{1}$, A.F. Volkov$^{1,2}$ and K.B. Efetov$^{1,3}$.}

\begin{abstract}
We analyze collective modes in two-band superconductors in the
dirty limit. It is shown that these modes exist at all
temperatures $T$ below $T_{c}$ provided the frequency of the modes
is higher than the inelastic scattering rate and lower than the
energy gaps $\Delta _{a,b}$. At low temperatures these modes are
related to counterphase oscillations of the condensate currents in
each band. The spectrum of the collective oscillations is similar
to the spectrum of the Josephson ''plasma''\ modes in a tunnel
Josephson junction but the velocity of the mode propagation in the
case under consideration is much lower. At higher temperatures
($\Delta _{b}<T<T_{c}$) the spectrum consists of two branches. One
of them is gapless (sound-like) and the second one has a
threshhold that depends on coupling between the bands. We
formulate the conditions under which both types of collective
modes can exist. The spectrum of the collective modes can be
determined by measuring the I-V characteristics of a Josephson
junction in a way as it was done by Carlson and Goldman \cite{CG}.
\end{abstract}

\maketitle

\address{$^{(1)}$ Theoretische Physik III,\\
 Ruhr-Universit\"{a}t Bochum, D-44780 Bochum, Germany\\
$^{(2)}$ Institute for Radioengineering and Electronics of Russian Academy of\\
 Sciences,11-7 Mokhovaya str., Moscow 125009, Russia\\
$^{(3)}$ L. D. Landau Institute for Theoretical Physics RAS,
119334 Moscow,
 Russia}

\bigskip

\section{Introduction}

The conventional BCS theory has been developed for a single-band metal with
an attractive interaction between electrons with opposite spins and momenta
(s-wave, singlet pairing) \cite{Schrieffer} and describes well most low-$%
T_{c}$ superconductors. The universality of the description of the
superconductors is a consequence of an assumption about a simple shape of
the Fermi surface.

However, some superconducting materials have a rather complicated band
structure. For example, there is a consensus that the recently discovered
new superconductor $MgB_{2}$ ($T_{c}\approx 40$ $K$) is a two-band
superconductor \cite{Nagamatsu}. In contrast to conventional BCS
superconductors, two-band superconductors may have two different order
parameters $\Delta _{a,b}$ (we label the bands by subscripts $a$ and $b$)
and an additional degree of freedom - the phase difference between the order
parameters: $\varphi =\chi _{a}-\chi _{b}$. This is a new variable that has
to be accounted for in the proper theory of such a superconductivity.
Naturally, one can expect new phenomena in two-band superconductors related
to this new degree of freedom.

One of the examples of this kind are $\varphi -$phase solitons predicted in %
\cite{Tanaka}. On the basis of the Ginzburg-Landau (GL) functional it was
shown that these solitons are described by the sine-Gordon equation. A
Gedanken-experiment that allows the observation of these solitons was
suggested in \cite{G+V}. The authors considered a two-band superconductor in
contact with a normal metal N. Using a version of the time-dependent GL
equation, they analyzed a current through the S/N interface and showed that
under certain conditions the phase solitons may be created in the
superconductor. Equation describing dynamics of these solitons is similar to
the one for a dissipative Josephson junction. In Ref. \cite{Babaev} an
analogy between the Ginzburg-Landau functional for two-band superconductors
and for some models in the particle physics (an extended version of
Faddeev's nonlinear $\sigma $ model) was used and, on this basis,
topologically different vortices in two-gap superconductors have been
predicted. In Ref.\cite{Nicol} the critical current in these superconductors
was calculated.

Another effect that can arise in two-band superconductors is weakly damped
oscillations of the phase difference $\varphi $ of the order parameters or,
in other words, collective modes (CMs) related to oscillations of the phase $%
\varphi $ in space and time. It is known that in single-band superconductors
CMs can exist only at temperatures close to the critical temperature $T_{c}$
(see reviews \cite{A+Vrev,SchoenRev}). In these modes the condensate current
$j_{S}$ and the quasiparticle current $j_{q}$ oscillate. Because the
variation of the total current density $\delta j$ should be zero due to the
quasineutrality condition ($\delta j=\delta j_{S}+\delta j_{q}=0$), the
oscillations of $\delta j_{S}$ are accompanied by counter-phase oscillations
of the quasiparticle current $\delta j_{q}\approx \sigma E$. The phase of
the order parameter also oscillates but the amplitude of the order parameter
$\Delta $ remains constant. These phase modes have an acoustic spectrum $%
\omega \sim kv_{cm}$ and exist in a sub-gap region ($\Delta ^{2}/T<\omega
<<\Delta ,1/\tau $; where $\tau $ is the elastic scattering time) in impure
superconductors. The CMs have been observed by Carlson and Goldman \cite{CG}
and explained theoretically in \cite{A+V,Schoen+Schmid}.

It is clear that CMs cannot exist in conventional superconductors at zero
temperature because in this limit only one degree of freedom, namely, the
condensate current $j_{S}$ exists. Any oscillations of the current density $%
j_{S}$ with not very high frequencies would lead to violation of the charge
neutrality. In contrast, in two-band superconductors, even at zero
temperature $T$ there are two degrees of freedom: the condensate currents $%
j_{Sa,b}$ in each band that can oscillate in counterphase such that the
total current density remains constant: $\delta j_{S}(r,t)=\delta
j_{Sa}(r,t)+\delta j_{Sb}(r,t)=0$. This is similar to what happens in
layered superconductors \cite{Layered} where the condensate currents in
different layers oscillate in counterphase. The CMs in the two-band
superconductors at zero (or low) temperature are related to oscillations of
the phase difference $\varphi .$ These CMs are similar to the Leggett mode
that can be excited in superfluid $He^{3}$ \cite{Leggett}.

Theoretically, the Leggett-type CMs in two-band superconductors at zero
temperature were studied in \cite{Sharapov}, and the influence of the CMs on
the Josephson effect in S$_{1}$/I/S$_{2}$ was analyzed in \cite{Agterberg}
(here S$_{1,2}$ is a single- and two-band superconductor). This Leggett-type
CM has a spectrum

\begin{equation}
\Omega ^{2}=\Omega _{0}^{2}+v_{L}^{2}k^{2},  \label{LegSpectrum}
\end{equation}%
which is typical for the Josephson tunnel junction, where $\Omega _{0}$ is a
threshold frequency, $v_{L}$ is a velocity of this CM (see Sec. III). The
similarity between the Josephson ``plasma waves''\ and the Leggett-type CM
in two-band superconductors is quite natural because the coupling between
different bands in two-band superconductors looks like the Josephson
coupling between superconductors in a Josephson tunnel junction.

According to estimates carried out in Ref. \cite{Sharapov}, the energy
corresponding to the threshold frequency $\Omega _{0}$ in $MgB_{2}$ is
higher than the smaller superconducting gap. This corresponds to a strong
damping of the CMs in this two-band superconductor. Therefore it would be of
interest to investigate under what conditions the gapless Carlson-Goldman
CMs can propagate in two-band superconductors and this is the subject of the
present paper.

The paper is organized as follows. In Sec.II we formulate a model of
two-band superconductors that will be used in calculations and present
microscopic equations for quasiclassical Green's functions. These equations
determine the spatial and temporal behavior of the retarded (advanced)
Green's functions $\hat{g}_{i}^{R(A)}(t,t^{\prime };r),$ as well as of the
Keldysh function $\hat{g}_{i}^{K}(t,t^{\prime };r).$ The functions $\hat{g}%
_{i}^{R(A)}$ and $\hat{g}_{i}^{K}$ are matrices in the particle-hole
(Gor'kov-Nambu) space. Using these equations we find a linear response $%
\delta \hat{g}_{i}^{R(A)}$ and $\delta \hat{g}_{i}^{K}$ to small
perturbations of the electric field and condensate velocities in both the
bands. This give us possibility to find the spectrum of the CMs at arbitrary
temperatures. Note that the spectrum of CMs cannot be obtained from the
generalized GL equations used in \cite{G+V}. In Sec.IV we analyze a method
that may enable one to observe the CMs.

\section{Model and Basic Equations}

We start our discussion considering a simple model of a two-band
superconductor and deriving microscopic equations for quasiclassical Green's
functions. Using these equations we can obtain equations for macroscopic
quantities and calculate the spectrum of CMs. We restrict ourselves with the
dirty limit assuming that the elastic impurity scattering time $\tau $ is
sufficiently small: $\Delta _{a,b}\tau <<1$. In the equilibrium case these
equations can be reduced to an Usadel-like equation used in \cite{Kosh+Gol}.

The Hamiltonian of the considered two-band superconductor has the form (see
for example \cite{Kosh+Var+Vin})

\begin{equation}
\mathit{H}=\mathit{H}_{a}+\mathit{H}_{b}-\sum_{\{p,q;i,k\}}\{V_{i,k}(q)\psi
_{i,p+q}^{\dagger }\psi _{i,-p}^{\dagger }\psi _{k,-p^{\prime }}\psi
_{k,p^{\prime }+q}+\mathrm{c.c.}\}\;,  \label{HamA}
\end{equation}%
with%
\begin{equation}
\mathit{H}_{a}=\sum_{\{p,q\}}\{\psi _{ap}^{\dagger
}v_{a}(p-p_{a})\psi _{ap}^{\dagger }+\psi _{a(p+q)}^{\dagger
}[V_{a,imp}(q)+V_{a}(q)]\psi _{ap}\} \label{Hama}
\end{equation}%
where the third term in Eq.(\ref{HamA}) describing the
electron-electron interaction leads to the superconductivity;
$p,q$ are momenta (strictly speaking, we must assign to $p$ also a
spin index $\sigma $ but we omit it
here for the sake of brevity) and the indices $\{i,k\}$ numerate the bands $%
\{a,b\}$. The first two terms are one-particle Hamiltonians for each band
that include the kinetic energy $\varsigma _{a,b}(p)=v_{a,b}(p-p_{a,b})$
counted from the Fermi level and the terms describing an elastic impurity
scattering. Beside a short-range potential $V_{a,imp}$ due to impurities, $H_{a}$
contains also a long-range self-consistent potential $V_{a}$ due to
coulomb interaction. The Hamiltonian for the $b$ band, $H_{b}$%
, is obtained from $H_{a}$\ by replacing subindeces $%
a\rightarrow b$\textbf{. }We neglect the interband impurity scattering
(arguments supporting this assumption have been given in \cite{Mazin02}).

The derivation of the equations for the quasiclassical Green's functions is
carried out in a standard way \cite{L+O,Kopnin}.\ However, the presence of
the two bands makes the situation more complicated. In order to avoid
unnecessary techincal difficulties, we make several assumptions: a) as in
the BCS theory we use the mean-field approximation representing the product
of four $\psi $ operators in the form $\Delta _{k}\psi _{k,-p^{\prime
}}^{\dagger }\psi _{k,p^{\prime }+q}^{\dagger }+\Delta _{i}\psi
_{i,-p^{\prime }}^{\dagger }\psi _{i,p^{\prime }+q}^{\dagger };$ b) we
neglect a change in the single-electron spectrum due to a possible tunneling
between the bands because the strong impurity scattering assumed here
destroys such a change, c) we neglect terms corresponding to pairing of
electrons from different bands $\langle \psi _{i,-p^{\prime }}^{\dagger
}\psi _{k,p^{\prime }+q}^{\dagger }\rangle $ (such a pairing was taken into
account in \cite{Ef+Lar} where the possibility of triplet pairing in a clean
layered superconductor was analyzed). With these assumptions used also in
previous works, we can derive a microscopic equation for the matrix
quasiclassical Green's functions $\check{g}$ in the same way as for a
one-band superconductor \cite{L+O,Kopnin}. This equation in the dirty limit (%
$\tau \Delta _{a,b}<<1$) has the standard form

\begin{equation}
-iD_{a}\nabla (\check{g}\nabla \check{g})_{a}+i[\check{\tau}_{3}\partial
\check{g}_{a}/\partial t+\partial \check{g}_{a}/\partial t^{\prime }\check{%
\tau}_{3}]+[\check{\Delta}_{a},\check{g}_{a}]+s_{a}[\check{\Delta}_{b},%
\check{g}_{a}]-e\left[ V(t)\check{g}-\check{g}V(t^{\prime })\right] =0;,
\label{Usadel}
\end{equation}%
where $D_{a}=(v_{F}l)_{a}/3$ is the diffusion coefficient in the $a$ band, $%
\check{g}_{a}(r;t,t^{\prime })$ is a 4$\times $4 matrix depending on the
coordinate $r$ and two times $t$ and $t^{\prime }$. The elements of this
matrix are the retarded (advanced) matrix Green's functions $\hat{g}^{R(A)}$
(elements (11) and (22)) and the matrix Keldysh function $\hat{g}^{K}$%
(element (12)). The parameter $s_{a}=V_{ab}/V_{b}$ determines the strength
of the coupling between superconducting pairing in the $a$ and $b$ bands
(this type of pairing for the two-band superconductors was suggested earlier
in \cite{Suhl}). The same equation as Eq. (\ref{Usadel}) is valid for the $b$
band provided the subscripts are exchanged, $a\leftrightarrows b$.

Eq.(\ref{Usadel}) is supplemented by the normalization condition

\begin{equation}
\check{g}_{a}(t,t_{1})\circ \check{g}_{a}(t_{1},t^{\prime })=\delta
(t-t^{\prime })  \label{NormCond}
\end{equation}
and the self-consistency equation

\begin{equation}
\hat{\Delta}_{a,b}=\lambda _{a,b}\hat{f}_{a,b}(t,t;r)  \label{SelfCon}
\end{equation}%
where $\lambda _{a,b}=(V\nu )_{a,b}$ and $\nu _{a,b}$ are the coupling
constant and the density of states in each band. If we wrote down Eq.(\ref%
{Usadel}) for the retarded (advanced) Green's functions in the Matsubara
representation, we would obtain a generalized Usadel equation. Such an
equation was used by Koshelev and Golubov \cite{Kosh+Gol}. Note that our
definition of the order parameter differs from the one $\left( \Delta
_{a}\right) _{KG}$ used in Ref. \cite{Kosh+Gol} and the correspondence
between both the definitions is given by $(\Delta _{a})_{KG}=\Delta
_{a}+s_{a}\Delta _{b}$.

The current density $j_{a,b}$ in each band is expressed in terms of the
Keldysh matrix as

\begin{equation}
\mathbf{j}_{a,b}(t,r)=(\pi /4)\sigma _{a,b}Tr\{\hat{\tau}_{3}[\hat{g}%
^{R}(t,t_{1};r)\nabla \hat{g}^{K}(t_{1},t;r)+\hat{g}^{K}(t,t_{1};r)\nabla
\hat{g}^{A}(t_{1},t;r)]_{a,b}\}  \label{Current}
\end{equation}

Our aim is to find the response $\delta \check{g}_{a,b}$ of the system to a
perturbation of the electric potential $V(r,t)$ and $\nabla \chi _{a,b}$. To
be more precise, we are interested in the response to perturbations of the
gauge-invariant potential $\mu _{a,b}$ and the condensate momentum $Q_{a,b}$

\begin{equation}
\mu _{a,b}=eV+(1/2)\partial \chi _{a,b}/\partial t;\text{ \ }\mathbf{Q}%
_{a,b}=(1/2)[\nabla \chi _{a,b}-(2\pi /\Phi _{0})\mathbf{A]}  \label{muQ}
\end{equation}%
where\ $\Phi _{0}=hc/2e$ is the magnetic flux quantum. These are the
responses that enter physical quantities. One more quantity $s_{a}[\check{%
\Delta}_{b},\check{g}_{a}]$ will also be considered as a perturbation with $%
\check{\Delta}_{b}$ and $\check{g}_{a}$ equal to their equilibrium values.

In equilibrium the Keldysh function $\hat{g}_{a,b}^{K}(\epsilon )$ equals

\begin{equation}
\hat{g}_{a,b}^{K}(\epsilon )=(\hat{g}^{R}(\epsilon )-\hat{g}^{A}(\epsilon
))_{a,b}\tanh (\epsilon \beta )  \label{gEquil}
\end{equation}%
where $\hat{g}_{a,b}^{R(A)}(\epsilon )=[\hat{\tau}_{3}g_{a,b}(\epsilon )+i%
\hat{\tau}_{2}f_{a,b}(\epsilon )]^{R(A)}$, $f_{a,b}^{R(A)}(\epsilon )=\Delta
_{a,b}/\xi _{a,b}^{R(A)}=(\Delta _{a,b}/\epsilon )g_{a,b}^{R(A)}(\epsilon ),$
$\xi _{a,b}^{R(A)}(\epsilon )=\sqrt{(\epsilon \pm i0)^{2}-\Delta _{a,b}^{2}}$
and $\beta =\left( 1/2T\right) .$ As usual, the quasiclassical functions $g$
and $f$ stand for the normal and condensate quasiclassical Green functions
and $\tau _{i}$ are the Pauli matrices in Gorkov-Nambu space. The method of
solution we will use is similar to that presented in \cite{A+Vrev}.

First, we single out the phase $\chi _{a,b}$ using the transformation $%
\check{g}_{a,b}=(\check{U}\check{g}_{new}\check{U}^{\dagger })_{a,b},$ where
$\check{U}_{a,b}=\exp (i\check{\tau}_{3}\chi _{a,b}/2).$ Then Eq.(\ref%
{Usadel}) for the new matrix $\check{g}_{new}$ acquires the form (we drop
the subindex $new$)

\begin{equation}
-iD_{a}\nabla (\check{g}\nabla \check{g})_{a}+i[\check{\tau}_{3}\frac{%
\partial \check{g}_{a}}{\partial t}+\frac{\partial \check{g}_{a}}{\partial
t^{\prime }}\check{\tau}_{3}]+[\check{\Delta}_{a},\check{g}_{a}]+s_{a}[%
\check{\Delta}_{b},\check{g}_{a}]-\left[ \mu (t)\check{g}-\check{g}\mu
(t^{\prime })\right] _{a}+D_{a}\mathbf{\nabla \cdot Q}_{a}\check{g}_{a}[%
\check{\tau}_{3},\check{g}_{a}]+iD_{a}\mathbf{Q}_{a}^{2}[\check{\tau}_{3},%
\check{g}_{a}\check{\tau}_{3}\check{g}_{a}]=0,  \label{NewUsad}
\end{equation}

After that we linearize Eq. (\ref{NewUsad}) \ with respect to perturbations $%
\delta \check{g}(t,t^{\prime };r)=\check{g}-\check{g}_{eq}$ and make the
Fourier transformations

\begin{equation}
\delta \check{g}(\epsilon ,\epsilon ^{\prime },\mathbf{k})=\int dtdt^{\prime
}\exp (i\epsilon t-i\epsilon ^{\prime }t^{\prime })\delta \check{g}%
(t,t^{\prime },\mathbf{k})
\end{equation}
where $\delta \check{g}(t,t^{\prime },\mathbf{r})\sim \delta \check{g}%
(t,t^{\prime },\mathbf{k})\exp (i\mathbf{kr})$. We represent the
perturbations $\mu (t,r)$ and $\mathbf{Q}(t,r)$ in the form: $\mu (t,r)\sim
\mu (\Omega ,\mathbf{k})\exp (i\mathbf{kr}-i\Omega t);$ $\mathbf{Q}(t,r)\sim
\mathbf{Q}(\Omega ,\mathbf{k})\exp (i\mathbf{kr}-i\Omega t)$.

Writing down equations for elements (11) and (22), i.e., for the matrices $%
\hat{g}^{R}$ and $\hat{g}^{A},$ we can obtain the expressions for the
perturbations $\delta \hat{g}^{R}(\epsilon ,\epsilon ^{\prime })$ of the
retarded Green's functions

\begin{equation}
\delta \hat{g}_{a}^{R}(\epsilon ,\epsilon ^{\prime })=\frac{1}{M_{a}^{R}}%
\{s_{a}(\hat{\Delta}_{b}-\hat{g}_{a\epsilon }^{R}\hat{\Delta}_{b}\hat{g}%
_{a\epsilon ^{\prime }}^{R})+\mu _{a}(\hat{g}_{a\epsilon }^{R}\hat{g}%
_{a\epsilon ^{\prime }}^{R}-1)-iD_{a}\mathbf{kQ}_{a}(\hat{\tau}_{3}\hat{g}%
_{a\epsilon }^{R}-\hat{g}_{a\epsilon ^{\prime }}^{R}\hat{\tau}_{3})\}
\label{DeltaGR}
\end{equation}%
where $M_{a}^{R}(\epsilon ,\epsilon ^{\prime })=(\xi _{\epsilon }^{R}+\xi
_{\epsilon ^{\prime }}^{R})_{a}+ik^{2}D_{a},(\xi _{\epsilon }^{R})_{a}=\xi
_{a}^{R}(\epsilon )$ and $\xi _{a}^{R}(\epsilon )$ is defined in Eq.(\ref%
{gEquil}).

The matrices of the perturbations $\delta \hat{g}_{b}^{R}(\epsilon ,\epsilon
^{\prime })$ and $\delta \hat{g}_{a}^{A}(\epsilon ,\epsilon ^{\prime })$ are
determined by Eq.(\ref{DeltaGR}) after the permutation of the subscripts: $%
a\rightarrow b$ and $R\rightarrow A$. Eq.(\ref{DeltaGR}) coincides with a
corresponding equation in \cite{A+Vrev} provided the limit $\Delta \tau <<1$
is taken and $s_{a}$ is set to zero: $s_{a}=0$.

In order to find the perturbation of the Keldysh function $\delta \hat{g}%
_{a}(\epsilon ,\epsilon ^{\prime })\equiv \delta \hat{g}_{a}^{K}(\epsilon
,\epsilon ^{\prime })$, we represent $\delta \hat{g}_{a}(\epsilon ,\epsilon
^{\prime })$ in the form of a sum of a regular $\delta \hat{g}_{reg}$ and
anomalous $\delta \hat{g}_{an}$ part

\begin{equation}
\delta \hat{g}_{a}(\epsilon ,\epsilon ^{\prime })=(\delta \hat{g}%
_{reg}(\epsilon ,\epsilon ^{\prime })+\delta \hat{g}_{an}(\epsilon ,\epsilon
^{\prime }))_{a}  \label{Delta G}
\end{equation}
where $\delta \hat{g}_{reg}(\epsilon ,\epsilon ^{\prime })=\delta \hat{g}%
^{R}(\epsilon ,\epsilon ^{^{\prime }})\tanh (\epsilon ^{\prime }\beta
)-\tanh (\epsilon \beta )\delta \hat{g}^{A}(\epsilon ,\epsilon ^{^{\prime
}}) $ (we drop the indeces $a,b$). The anomalous part is obtained in a way
similar to that in \cite{A+Vrev}. It has the form

\begin{equation}
(\delta \hat{g}_{an}(\epsilon ,\epsilon ^{\prime }))_{a}=\frac{\tanh
(\epsilon ^{\prime }\beta )-\tanh (\epsilon \beta )}{M_{a}}\{-s_{a}(\hat{%
\Delta}_{b}-\hat{g}_{a\epsilon }^{R}\hat{\Delta}_{b}\hat{g}_{a\epsilon
^{\prime }}^{A})+\mu _{a}(1-\hat{g}_{a\epsilon }^{R}\hat{g}_{a\epsilon
^{\prime }}^{A})-iD_{a}\mathbf{kQ}_{a}(\hat{g}_{a\epsilon }^{R}\hat{\tau}%
_{3}-\hat{\tau}_{3}\hat{g}_{a\epsilon ^{\prime }}^{A})\}  \label{DeltaGK}
\end{equation}%
where $M_{a}(\epsilon ,\epsilon ^{\prime })=(\xi _{\epsilon }^{R}+\xi
_{\epsilon ^{\prime }}^{A})_{a}+ik^{2}D_{a}.$ The energies $\epsilon
,\epsilon ^{\prime }$ in Eqs.(\ref{DeltaGR}-\ref{DeltaGK}) are equal to $%
\epsilon =\bar{\epsilon}+\Omega /2,$ $\epsilon ^{\prime }=\bar{\epsilon}%
-\Omega /2$, where $\bar{\epsilon}=(\epsilon +\epsilon ^{\prime })/2$ and $%
\Omega $ is the frequency of oscillations.

Having determined the perturbations $\delta \hat{g}_{a}^{R(A)}(\epsilon
,\epsilon ^{\prime })$ and $\delta \hat{g}_{a}(\epsilon ,\epsilon ^{\prime
}) $, we can readily derive equations for such macroscopic quantities as $%
\mu _{a,b}(t,r),\delta j_{a,b}(t,r),$ etc, in each band and obtain the
spectrum of the CMs.

\bigskip

\section{Macroscopic Quantities. Spectrum of Oscillations.}

As follows from its definition, the condensate momentum $\mathbf{Q}_{a,b}$
obeys the equations

\begin{equation}
\partial \mathbf{Q}_{a,b}/\partial t=e\mathbf{E}+\nabla \mu _{a,b}
\label{EqQ}
\end{equation}%
In order to obtain an equation for $\mu _{a,b}$, we can use the
self-consistency equation (\ref{SelfCon}) written for the phases $\chi
_{a,b} $. This means that terms proportional to $\hat{\tau}_{1}$\ in Eq.(\ref%
{SelfCon}) should be equal to zero. The variation of the current density is
found from Eq.(\ref{Current}).

We consider first the case of low temperatures: $T<<\Delta _{a,b}$.

a) $T<<\Delta _{a,b}$. In this case the main contribution is due to the
regular part $\delta \hat{g}_{reg}(\epsilon ,\epsilon ^{\prime })$. The
anomalous part gives small corrections of the order $\Omega /\Delta _{a,b}$
because we assume that $\Omega /\Delta _{a,b}<<1$.

Let us obtain an equation for $\mu _{a,b}$ using Eq.(\ref{SelfCon}).
Calculating the contribution from the regular part and setting $\epsilon
=i\omega +\Omega /2$ and $\epsilon ^{\prime }=i\omega -\Omega /2$ we can
transform the integration over $\bar{\epsilon}$ into a sum over the
Matsubara frequencies:
\[
\int d\bar{\epsilon}[\delta \hat{g}^{R}(\epsilon ,\epsilon ^{^{\prime
}})\tanh (\epsilon ^{\prime }\beta )-\tanh (\epsilon \beta )\delta \hat{g}%
^{A}(\epsilon ,\epsilon ^{^{\prime }})]=(2\pi i)(2T)\sum_{\omega }\delta
\hat{g}^{R}(\epsilon ,\epsilon ^{^{\prime }})
\]%
Substituting the perturbations $\delta \hat{g}$ into Eq. (\ref{SelfCon}) we
get

\begin{equation}
-\epsilon _{0}^{2}\sin \varphi -\bar{\nu}_{a}\partial \mu _{a}/\partial
t+\Delta _{a}(\bar{\nu}_{a}D_{a})\nabla \mathbf{Q}_{a}=0  \label{EqMu}
\end{equation}%
where $\epsilon _{0}^{2}=2(V_{ab}/V_{a}V_{b}\nu )\Delta _{a}\Delta _{b},\bar{%
\nu}_{a}=\nu _{a}/(\nu _{a}+\nu _{b})$ is the normalized density-of-states
in the $a$ band, $\nu =\nu _{a}+\nu _{b}$, and $\varphi =\chi _{a}-\chi
_{b}. $ The same equation with the interchange of indices $a\leftrightarrows
b$ is valid for the $b$ band (one has to keep in mind that the sign in front
of $\sin \varphi $ changes as a result of this interchange).

Eq. (\ref{EqMu}) is the continuity equation for the charge of Cooper pairs $%
q_{a}\sim \nu _{a}\mu _{a}$ in the $a$ band because the third term is
proportional to the divergence of the condensate current $j_{a}$. The first
term in\ Eq. (\ref{EqMu}) describes a Josephson-like coupling between the
bands and may be considered as a drain ($\varphi >0$) or source ($\varphi <0$%
) of Cooper pairs in the $a$ band. Eq.(\ref{EqMu}) is valid for frequencies
exceeding the effective relaxation time $\tau _{imb}$ for a charge
imbalance, $\Omega >>1/\tau _{imb}$. This relaxation time is determined, in
particular, by the electron-phonon and electron-electron inelastic
scattering. If charge imbalance relaxation processes are taken into account,
the term $\partial \mu _{a}/\partial t$ in Eq.(\ref{EqMu}) should be
replaced by ($\partial /\partial t+\gamma $), where $\gamma ^{-1}=\tau
_{imb} $ \cite{Tinkham,A+Vrev,SchoenRev}.

At low temperatures the current density $j_{a}$ coincides in the main
approximation with the condensate current and equals

\begin{equation}
\mathbf{j}_{a}=\pi \sigma _{a}\Delta _{a}\mathbf{Q}_{a}/e  \label{ScurrentT0}
\end{equation}

This expression and Eqs.(\ref{EqQ}), (\ref{EqMu}) describe the system at low
temperatures. One should add also the charge neutrality condition

\begin{equation}
\delta \mathbf{j=}\delta \mathbf{(j}_{a}+\mathbf{j}_{b})=0.
\label{QuasiNeutr}
\end{equation}

Then, we can exclude all the variables except $\varphi $ and
obtain the equation for the phase difference

\begin{equation}
\Omega _{0}^{2}\sin \varphi +(\partial /\partial t+\gamma )\partial \varphi
/\partial t-v_{cm}^{2}\nabla ^{2}\varphi =0  \label{SineGor}
\end{equation}
where $\Omega _{0}^{2}=2\epsilon _{0}^{2}/(\bar{\nu}_{a}\bar{\nu}_{b}),$ $%
\gamma =1/\tau _{imb}$ is a damping rate and

\begin{equation}
v_{cm}=[\pi \Delta _{a}\Delta _{b}\frac{D_{a}D_{b}}{D_{a}\bar{\nu}_{a}\Delta
_{a}+D_{b}\bar{\nu}_{b}\Delta _{b}}]^{1/2}  \label{Velocity}
\end{equation}
is velocity of the CMs or the limiting velocity of phase solitons in the two
band superconductors.

Eq.(\ref{SineGor}) is similar to the sine-Gordon equation for a tunnel
Josephson junction but the velocity $v_{cm}$ is much smaller than the
corresponding velocity of the Swihart waves in a Josephson junction. By the
order of magnitude the velocity $v_{cm}$ is equal to the velocity ($\sim
\sqrt{D\Delta }$) of the Carlson-Goldman CM in ordinary superconductors.
However, in contrast to the Carlson-Goldman CMs in single-band
superconductors that are weakly damped only near $T_{c}$, the CM described
by Eq.(\ref{SineGor}) exist at low temperatures. In these modes, condensate
in the $a$ band oscillates with respect to the condensate in the $b$ band
and these oscillations are accompanied by oscillations of the phase
difference $\varphi .$ The spectrum of the small amplitude oscillations is
given by

\begin{equation}
\Omega ^{2}=\Omega _{0}^{2}+k^{2}v_{cm}^{2}.  \label{Spectrum0}
\end{equation}

\begin{figure}[tbp]
\par
\begin{center}
\includegraphics[width=0.95\textwidth]{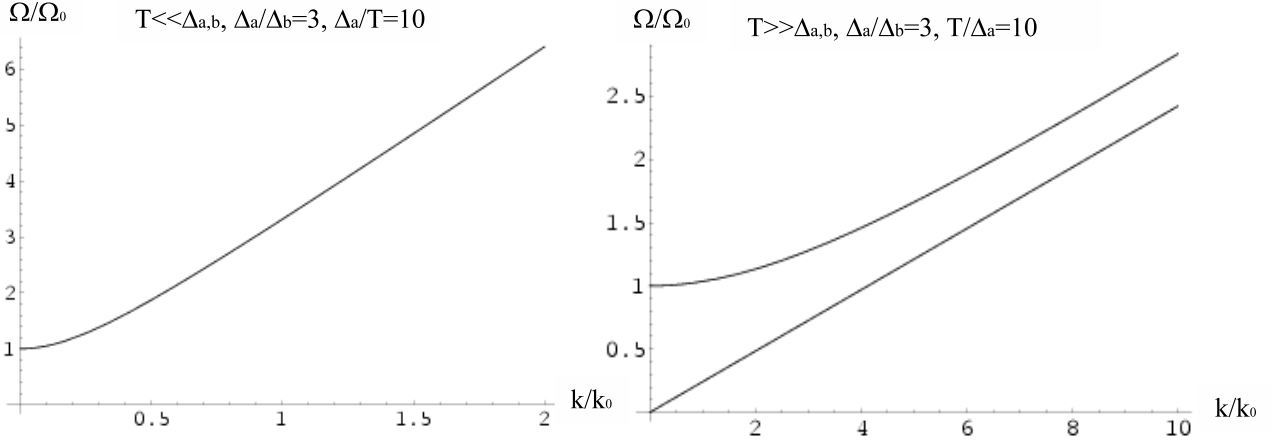}
\end{center}
\par
.
\caption{a) Calculated CM spectrum for the case of low temperatures ($%
T<<\Delta _{a,b}$). There is a gap $\Omega_0$ in the CM spectrum. ($%
k_{o}=\Omega_{0}/v_{CM}$) b) The CM spectrum for high temperatures ($%
T>>\Delta _{a,b}$). There are two branches of the CMs: the sound-like mode
and the mode with a gap (analogous to the Leggett mode). The sound-like mode
exists for frequencies greater than inverse energy relaxation time: $\Omega
> \protect\gamma$.}
\end{figure}

The threshold frequency $\Omega _{0}$ is analogous to the Josephson ``plasma
frequency''. As we assume that $\Omega $ is less than $\Delta _{a,b}$, our
consideration is valid provided $V_{ab}(\nu _{a}+\nu _{b})/2V_{a}V_{b}\nu
_{a}\nu _{b}<\Delta _{b}/\Delta _{a}.$ This type of CMs that is analogous to
the Leggett mode in $He^{3}$ was studied theoretically in Refs.\cite%
{Sharapov,Agterberg}. The dependence of $\ \Omega (k)$ is depicted in Fig.1a
for certain parameters of the model.

Now we consider the limit of temperatures close to $T_{c}$.

b) $\Delta _{a,b}<<$\ $T$. The calculations in this case become more
combersome because the anomalous Green's function $\delta \hat{g}%
_{an,a}(\epsilon ,\epsilon ^{\prime })$, Eq.(\ref{DeltaGK}), also gives an
essential contribution. In this case the currents $j_{a,b}$ are equal to

\begin{equation}
\mathbf{j}_{a,b}=\sigma _{a,b}[(\pi \Delta _{a,b}^{2}/2T)\mathbf{Q}_{a,b}/e+%
\mathbf{E]}  \label{CurrentTc}
\end{equation}%
where the first term is the supercurrent and the second term is the
quasiparticle current $j_{q}=\sigma E.$ In the main approximation this
current originates from the anomalous part $\delta \hat{g}_{an,a}(\epsilon
,\epsilon ^{\prime }),$ Eq.(\ref{DeltaGK}).

The equation for $\mu _{a}$ acquires the form

\begin{equation}
\tilde{\epsilon}_{0}^{2}\sin \varphi +(\bar{\nu}p)_{a}(\partial /\partial
t+\gamma )\mu _{a}-(\bar{\nu}p)_{a}v_{a}^{2}\mathbf{\nabla Q}_{a}=0
\label{EqMuT}
\end{equation}
where $\tilde{\epsilon}_{0}^{2}=(4T/\pi \Delta )\epsilon _{0}^{2}$, $\Delta
=\Delta _{a}+\Delta _{b},p_{a}=\bar{\nu}_{a}\Delta _{a}/(\Delta _{a}+\Delta
_{b})$ and $v_{a}=\sqrt{2D_{a}\Delta _{a}}$ is the velocity of the
Carlson-Goldman mode in the $a$-band. In the limit $\{\gamma ,\Delta
_{a,b}^{2}/T\}<\Omega <\Delta _{a,b}$ the equation for $Q_{a,b}$ is reduced
in the main approximation to $\partial \mathbf{Q}_{a,b}/\partial t\approx
\mathbf{\nabla }\mu _{a,b}$. The spectrum of the CMs consists of two
branches determined by the roots of equation

\begin{equation}
(\Omega ^{2}-k^{2}v_{a}^{2})(\Omega ^{2}-k^{2}v_{b}^{2})=2\tilde{\epsilon}%
_{0}^{2}[(\Omega ^{2}-k^{2}v_{a}^{2})/p_{b}+(\Omega
^{2}-k^{2}v_{b}^{2})/p_{a})]  \label{SpectrumTc}
\end{equation}

In the long-wave limit these branches are described by the expressions

\begin{equation}
\Omega _{1}^{2}=2\tilde{\epsilon}%
_{0}^{2}(1/p_{a}+1/p_{b})+k^{2}(v_{a}^{2}p_{b}+v_{b}^{2}p_{a})/(p_{a}+p_{b})
\label{Omega1}
\end{equation}
and

\begin{equation}
\Omega _{2}^{2}=k^{2}(v_{a}^{2}p_{a}+v_{b}^{2}p_{b})/(p_{a}+p_{b})
\label{Omega2}
\end{equation}%
Therefore one branch of the CMs has a sound-like spectrum ($\Omega
_{2}$) and another one a Josephson-like spectrum ($\Omega _{1}$).
If $\nu _{a}>>\nu _{b}$ and $\Delta _{a}>>\Delta _{b}$ (this limit
corresponds to the two-band superconductor $MgB_{2}$), one has
$\Omega _{2}^{2}=k^{2}v_{a}^{2}$, that is, the low-frequency mode
coincides with the Carlson-Goldman mode in the band with a higher
gap. This low frequency mode may be excited in such two-band
superconductor as $MgB_{2}$. One can show that if $\hbar \Omega
<\Delta _{a}^{2}/T$, then the sound-like branch of the CMs is
strongly damped. Therefore at low temperatures, the ''soft'' mode
can hardly exist. Below we clarify this point in more detail. The
form of the spectrum at high temperatures is shown in Fig.1b.
Consider now the case of intermediate temperatures.

c) $\Delta _{b}<T<\Delta _{a}$.

The analysis given above shows that the equation for the potential $\mu $
can be written in limiting cases as
\begin{equation}
\mp \epsilon _{0}^{2}\sin \varphi -(\bar{\nu}r)_{a,b}[\partial \mu /\partial
t-\tilde{v}^{2}]_{a,b}\nabla \mathbf{Q}_{a,b}=0  \label{MuGeneral}
\end{equation}%
where the coefficients \ $r_{a,b}$ and velocities $\tilde{v}_{a,b}^{2}$ are
equal to

\begin{equation}
r_{a,b}=\left\{
\begin{array}{l}
1, \\
\pi \Delta _{a,b}/4T,%
\end{array}
\right. \text{ \ }\tilde{v}_{a,b}^{2}=v_{a,b}^{2}\left\{
\begin{array}{ll}
\pi /2, & T<<\Delta _{a,b} \\
1, & T>>\Delta _{a,b}%
\end{array}
\right.
\end{equation}
where $v_{a,b}^{2}=\sqrt{2(D\Delta )_{a,b}}$. The current density in each
band is given by

\begin{equation}
\mathbf{j}_{a,b}=(\sigma \rho )_{a,b}\Delta _{a,b}\mathbf{Q}_{a,b}+\tilde{%
\sigma}_{a,b}\mathbf{E}\text{ }  \label{CurrentInter}
\end{equation}
with limiting values of conductivities

\begin{equation}
(\sigma \rho )_{a,b}=\sigma _{a,b}\left\{
\begin{array}{l}
\pi , \\
\pi \Delta _{a,b}/2T,%
\end{array}
\right. \tilde{\sigma}_{a,b}=\sigma _{a,b}\left\{
\begin{array}{ll}
\exp (-\Delta _{a,b}/T), & T<<\Delta _{a,b} \\
1, & T>>\Delta _{a,b}%
\end{array}
\right.
\end{equation}

Qualitatively, dynamics of the CMs at any temperatures is described by Eqs.(%
\ref{EqQ},\ref{QuasiNeutr},\ref{MuGeneral},\ref{CurrentInter}). One can
exclude the varaibles $\mathbf{Q}_{a,b}$ and $\mathbf{E}$ and obtain the
equations

\begin{eqnarray}
\bar{\nu}_{a}A_{a}\mu _{a}-\bar{\nu}_{a}B_{b}\mu _{b} &=&-\epsilon
_{0}^{2}\varphi ,  \label{MuFi1} \\
\bar{\nu}_{b}A_{b}\mu _{b}-\bar{\nu}_{b}B_{a}\mu _{a} &=&\epsilon
_{0}^{2}\varphi ,  \label{MuFi2}
\end{eqnarray}%
where $A_{a}=r_{a}[-i\Omega +($\ $\tilde{v}_{a}k)^{2}(-i\Omega +\Omega _{b})/%
\mathit{M}];$ $B_{a}=($\ $\tilde{v}_{b}k)^{2}r_{b}\Omega _{a}/\mathit{M};$ $%
\mathit{M}=i\Omega (i\Omega -\Omega _{+});$ $\Omega _{a}=\rho _{a}\bar{\sigma%
}_{a}\Delta _{a};$ $\Omega _{+}=\Omega _{a}+\Omega _{b}$; $\bar{\sigma}%
_{a}=\sigma _{a}/(\tilde{\sigma}_{a}+\tilde{\sigma}_{b})$. Using the
relation $\mu _{a}-\mu _{b}=(1/2)\partial \varphi /\partial t$, we obtain
the dispersion relation

\begin{equation}
i\Omega \bar{\nu}_{a}\bar{\nu}_{b}(A_{a}A_{b}-B_{a}B_{b})=2\epsilon _{0}^{2}(%
\bar{\nu}_{a}A_{a}+\bar{\nu}_{b}A_{b}-\bar{\nu}_{b}B_{a}-\bar{\nu}_{a}B_{b})
\label{DisperGen}
\end{equation}

\begin{figure}[tbp]
\par
\begin{center}
\includegraphics[width=0.6\textwidth]{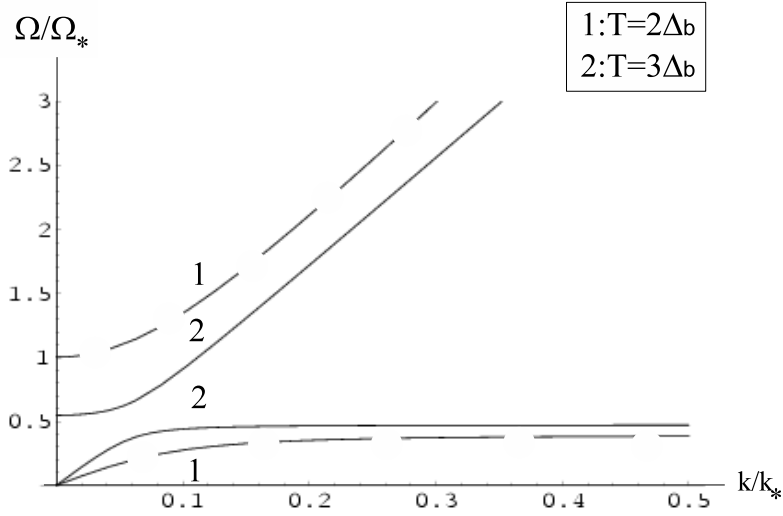}
\end{center}
\caption{CM spectrum at intermediate temperatures ($\Delta_{b}<T<\Delta_{a}$%
) for two cases: $\Omega>>\Omega_{a,b}$ (a) and $\Omega<<\Omega_{a,b}$ (b).
There are two branches of the CM spectrum in the both limiting cases: the
sound-like mode and the mode with a gap. \newline
The frequency and wave vector of CMs are normalized to $\Omega_{*}$ and $%
k_{*}$ equal to: a) $\Omega_{*}=E_{1}(T=8 \Delta_{b}), k_{*}=E_{1}/v_{1}$
and b) $\Omega_{*}=E_{1}(T=2 \Delta_{b}), k_{*}=E_{1}/v_{1}$. With
increasing temperature the two modes are getting closer to each other.}
\end{figure}

In the limiting cases $T<<\Delta _{a,b}$\ and $\Delta _{a,b}<<T$\ we obtain
the formulas for the CM spectrum given above (see Eqs.(\ref{Spectrum0},\ref%
{Omega1},\ref{Omega2})). The ''soft'', sound-like mode, which exists in the
temperature range $\Delta _{a,b}<<T<<T_{c},$\ is of a special interest
because another mode with a threshold frequency of the order of $\epsilon
_{0}/\hbar $\ may be strongly damped if $\epsilon _{0}>\Delta _{b}$\ (this
case seems to correspond to $MgB_{2}$\ \cite{Sharapov}).

The spectrum of this and other (plasma-like) mode may be obtained from Eq. (%
\ref{DisperGen}). Consider different limits of high and low
frequencies $\Omega $ with respect to the characteristic
frequencies $\Omega _{a,b}.$ These frequencies in the considered
temperature range ($\Delta _{b}<T<\Delta _{a}$) are equal to: $%
\Omega _{a}\cong \pi (\sigma _{a}/\sigma _{b})\Delta _{a},$ $\Omega
_{b}\cong \pi \Delta _{b}^{2}/2T$.

c1) Assume that the frequency of oscillations is low

\begin{equation}
\Omega <<\Omega _{a,b}
\end{equation}

In the considered temperature range this condition implies that $%
\Omega <<\Delta _{b}^{2}/T<<\Delta _{b}<<\Delta _{a}$\textbf{. }Then the
functions $A_{a,b}$\ and $B_{a,b}$\ acquire the form

\begin{equation}
A_{a,b}\approx (r_{a,b}/i\Omega )[\Omega ^{2}-(k\text{\ }\tilde{v}_{a,b})^{2}%
\frac{\Omega _{b,a}}{\Omega _{+}}]
\end{equation}

\begin{equation}
B_{a,b}\approx -(r_{b,a}/i\Omega )(k\tilde{v}_{b,a})^{2}\frac{\Omega _{a,b}}{%
\Omega _{+}}
\end{equation}

Substituting these expressions for $A_{a,b}$\ and $B_{a,b}$, we
obtain the spectrum of oscillations

\begin{equation}
\Omega ^{2}{}=2\epsilon _{1}^{2}+k^{2}v_{1}^{2}  \label{Roots}
\end{equation}%
where $\epsilon _{1}^{2}=\epsilon _{0}^{2}[(\bar{\nu}%
_{a}r_{a})^{-1}+(\bar{\nu}_{b}r_{b})^{-1}],v_{1}^{2}=(\tilde{v}%
_{a}^{2}\Omega _{b}+\tilde{v}_{b}^{2}\Omega _{a})/\Omega _{+},$
$r_{a}\cong 1,$ $r_{b}\cong \pi \Delta _{b}^{2}/4T$. We see that
in this case the spectrum has a threshold frequency $\sim $\
($8/\pi )\epsilon _{0}^{2}T/(\Delta _{b}\bar{\nu}_{b})$. The
sound-like mode is strongly damped in this limit.

c2) For large frequencies satisfying the condition

\begin{equation}
\Omega >>\Omega _{a,b}
\end{equation}
we can write the functions $A_{a,b}$ and $B_{a,b}$ in the form

\begin{equation}
A_{a,b}\approx (r_{a,b}/i\Omega )[\Omega ^{2}-(k\tilde{v}_{a,b})^{2}];%
\text{ \ }B_{a,b}\approx 0
\end{equation}

\bigskip The roots of Eq.(\ref{DisperGen}) are

\begin{equation}
\Omega _{1,2}^{2}{}=2\epsilon _{1}^{2}+\frac{1}{2}k^{2}(\tilde{v}_{a}^{2}+%
\tilde{v}_{b}^{2})\pm \{\epsilon _{1}^{4}+\frac{1}{4}k^{4}(\tilde{v}_{a}^{2}-%
\tilde{v}_{b}^{2})+\epsilon _{1}^{2}k^{2}(\tilde{v}_{a}^{2}+\tilde{v}%
_{b}^{2})-2\epsilon _{0}^{2}k^{2}(\frac{\tilde{v}_{a}^{2}}{(r\bar{\nu})_{b}}+%
\frac{\tilde{v}_{b}^{2}}{(r\bar{\nu})_{a}})\}^{1/2}  \label{Rts1}
\end{equation}%
In the long-wave limit we obtain for $\Omega _{1}^{2}$ and $\Omega
_{2}^{2}$

\begin{equation}
\Omega _{1}^{2}{}=2\epsilon _{1}^{2}+k^{2}v_{2}^{2};\text{ \ }\Omega
_{2}^{2}{}=k^{2}v_{3}^{2}  \label{Roots2}
\end{equation}%
where $v_{2}^{2}=(\tilde{v}_{a}^{2}(\bar{\nu}r)_{b}+\tilde{v}%
_{b}^{2}(\bar{\nu}r)_{a}/(\bar{\nu}_{a}r_{a}+\bar{\nu}_{b}r_{b}),$ $%
v_{3}^{2}=(\tilde{v}_{a}^{2}(\bar{\nu}r)_{a}+\tilde{v}_{b}^{2}(\bar{\nu}%
r)_{b})/(\bar{\nu}_{a}r_{a}+\bar{\nu}_{b}r_{b}).$ Thus, we have a
''hard'' mode ($\Omega _{1}{}$) and a ''soft'' mode ($\Omega
_{2}{} $). However, the considered limit of high frequencies
corresponds to a hypothetical case which is hardly realizable in
experiments. We assumed that the frequency $\Omega $ is higher than %
$\Omega _{a,b}$, but lower than $\Delta _{a,b}$. This means, in
particular, that the inequality $\Omega _{a}<\Delta _{b}$
 should be satisfied. This inequality can be presented in the form: $%
\sigma _{a}/\sigma _{b}<<(\Delta _{b}/\pi \Delta _{a})$; that is,
the conductivity of the $a$ band with the higher energy gap
$\Delta _{a}$ should be much less than the conductivity of the $b$
band.

In Fig.2 we plot the dispersion curves $\Omega (k)$ for different
temperatures determined by Eq.(\ref{Rts1}).

\section{Observation of collective modes in a Josephson junction}

In this section we analyze an experimental method that allows one to observe
the CMs and to determine the spectrum of these modes. The idea to identify
the CMs in a superconductor has been suggested by Carlson and Goldman \cite%
{CG,A+Vrev,SchoenRev} and also used in Ref.\cite{Agterberg}.

According to this idea one has to measure the I-V characteristics of a
tunnel Josephson junction in the presence of a bias voltage $V_{B}$ across
the junction and of a weak external magnetic field $H$. As in Ref.\cite%
{Agterberg}, we consider a tunnel Josephson junction $S_{1}/I/S_{2}$
consisting of two superconductors one of which is a two-band superconductor (%
$S_{2}$) and another one ($S_{1}$) is a single band superconductor. However,
unlike Ref.\cite{Agterberg} we take into account a magnetic field, which
allows one to determine the spatial dispersion of the CM spectrum and use
the microscopic approach to derive necessary equations. As in Ref. \cite%
{Agterberg}, we calculate the Josephson current $j_{J}$ taking into account
terms of high order in transparence $|T_{k}|^{2}.$ As is well known, in the
main approximation the Josephson current at a finite voltage $V_{B}$ and in
the presence of a magnetic field $H$ has the form of a traveling wave (see,
for example, \cite{Abrikosov,Kulik,Barone} )

\begin{equation}
j_{J}(x,t)=j_{J}\sin (\Omega _{V}t-K_{H}x).  \label{JosCurr1}
\end{equation}
where $\Omega _{V}=2eV_{B}/\hbar $ and $K_{H}$ is given by Eq.(\ref{A9}).
This current is injected into the superconductor $S_{2}$ and modifies Eqs.(%
\ref{EqMuT}) for the \ gauge-invariant potentials $\mu _{a,b}$ describing
the conservation of charge of the Cooper pairs in each band. We show that,
as in the Carlson and Goldman experiment \cite{CG}, a resonance occurs if
the frequency $\Omega $ and the wave vector $k$ coincide with $\Omega _{V}$
and $K_{H}$ respectively. As was noted in Ref.\cite{Agterberg}, this
resonance is analogous to the Fiske resonance in a Josephson tunnel junction
when the velocity of the traveling wave (\ref{JosCurr1}) $\Omega _{V}/K_{H}$
coincides with velocity of the Josephson ''plasma''\ waves.

In order to generalize Eqs.(\ref{EqMuT}) to the case of a Josephson
junction, we assume that the thickness of the $S_{2}$ electrode $d$ is
smaller than the London penetration depth $\lambda _{L}$ as well as the
quantity $k^{-1}$ and average Eq.(\ref{Usadel}) over the thickness $d$
taking into account the boundary conditions \cite{Zaitsev,K+L}

\begin{equation}
(\check{g}\partial \check{g}/\partial z)_{a}=(2R_{B}\sigma )_{a}^{-1}[\check{%
g}_{S,}\check{g}_{a}]  \label{BC}
\end{equation}
where $R_{Ba}$ and $\sigma _{a}$ are the interface resistance and
conductivity for electrons in the $a$ band in the normal state. The matrix
Green's function $\check{g}_{S}$ in the one-band superconductor is assumed
to describe a single band superconductor in equilibrium. The retarded
(advanced) Green's functions in the $S$-electrode have the standard form

\begin{equation}
\hat{g}_{S}^{R(A)}(\epsilon )=[\hat{\tau}_{3}g_{S}(\epsilon )+(i\hat{\tau}%
_{2}\cos \chi _{S}+i\hat{\tau}_{1}\sin \chi _{S})f_{S}(\epsilon )]^{R(A)}
\label{gR(A)}
\end{equation}%
with $f_{S}^{R(A)}(\epsilon )=\Delta _{S}/\xi _{S}^{R(A)}=(\Delta
_{S}/\epsilon )g_{S}^{R(A)}(\epsilon ),$ $\xi _{S}^{R(A)}(\epsilon )=\sqrt{%
(\epsilon \pm i0)^{2}-\Delta _{S}^{2}};$ $\chi _{S}$ is the phase in the $S$
electrode. This averaging over the thickness $d$ leads to the appearance in
Eq.(\ref{NewUsad})\ an additional term of the form $iD_{a}(2R_{B}\sigma
)_{a}^{-1}[\check{g}_{S,}\check{g}_{a}].$ The modified equation can be
solved in the same way as it was done before. For example, the retarded
function $\hat{g}_{a}^{R}$ is given again by Eq.(\ref{DeltaGR}) with an
additional term

\begin{equation}
i\tilde{E}_{Ja}(\hat{g}_{S}^{R}-\hat{g}_{a}^{R}\hat{g}_{S}^{R}\hat{g}%
_{a}^{R})  \label{AddTerm}
\end{equation}%
where the energy $\tilde{E}_{Ja}=D_{a}/(2R_{B}\sigma )_{a}d$ is related to
the interface transparency and to the Josephson coupling energy, $d$ is the
thickness of the two-band superconductor. This term results in a
corresponding modification of Eq.(\ref{MuGeneral}), that acquires the form

\begin{equation}
\mp \epsilon _{0}^{2}\sin \varphi -(\bar{\nu}s)_{a,b}\partial \mu
_{a,b}/\partial t+(\bar{\nu}D/\sigma )_{a,b}(\nabla \mathbf{j}%
_{a,b}-(j_{a,b}/d)\sin (\chi _{a,b}-\chi _{S}))=0  \label{MuJos}
\end{equation}%
where the current $j_{a,b}$ at low temperatures is given by Eq.(\ref%
{ScurrentT0}) and the Josephson critical current $j_{a,b}$ is equal to $%
j_{a,b}=(\tilde{\Delta}/eR_{B})_{a,b}$, where the parameter $\tilde{\Delta}$%
\ equals

\begin{equation}
\tilde{\Delta}_{a}=2\pi T\sum_{\omega =0}^{\infty }\frac{\Delta _{a}}{\sqrt{%
\omega ^{2}+\Delta _{a}^{2}}}\frac{\Delta _{S}}{\sqrt{\omega ^{2}+\Delta
_{S}^{2}}}  \label{EffDelta}
\end{equation}

At low temperatures $\tilde{\Delta}_{a}=\Delta _{S}\ln (c_{1}\Delta
_{a}/\Delta _{S})$\ if $\Delta _{S}<<\Delta _{a}$\ and $\tilde{\Delta}%
_{a}=\Delta _{a}\ln (c_{1}\Delta _{S}/\Delta _{a})$\ if $\Delta _{S}>>\Delta
_{a},$ where $c_{1}=2\int_{0}^{\infty }dxx\ln (2x)/(x^{2}+1)^{3/2}\approx
2.8.$

The phases $\chi _{a,b}$ may be represented in the form

\begin{equation}
\chi _{a,b}=\bar{\chi}\pm \varphi /2  \label{Phases_a,b}
\end{equation}
where $\bar{\chi}=(\chi _{a}+\chi _{b})/2.$

We assume that there are no $\varphi $-solitons in the system (according to
Ref.\cite{G+V} special conditions are needed to create such solitons) so
that in the main approximation one has: $\chi _{a}=\chi _{b};$ $\bar{\chi}%
-\chi _{S}=2eV_{B}t/\hbar -K_{H}x$ (see Appendix)$.$ Substituting these
expressions into Eq.(\ref{MuJos}), we obtain two equations

\begin{eqnarray}
-\epsilon _{0}^{2}\sin \varphi -\bar{\nu}_{a}\partial \mu _{a}/\partial t+(%
\bar{\nu}\alpha )_{a}\nabla \mathbf{Q}_{a} &=&E_{Ja}^{2}\sin (\Omega
_{V}t-K_{H}x),  \label{EqPhi1} \\
\epsilon _{0}^{2}\sin \varphi -\bar{\nu}_{b}\partial \mu _{b}/\partial t+(%
\bar{\nu}\alpha )_{b}\nabla \mathbf{Q}_{b} &=&E_{Jb}^{2}\sin (\Omega
_{V}t-K_{H}x),  \label{EqPhi2}
\end{eqnarray}
where $\alpha _{a}=\pi (\sigma \Delta )_{a}/e$ and $E_{Ja}^{2}=2\tilde{\Delta%
}_{a}\bar{\nu}_{a}\tilde{E}_{Ja}$. Taking into account Eqs.(\ref{EqQ}) for $%
\mathbf{Q}_{a,b}$, we can find the phase perturbation $\varphi $ due to the
''external forces'' $E_{Ja,b}^{2}\sin (\Omega _{V}t-K_{H}x)$

\begin{equation}
\varphi =2(E_{Ja}^{2}/\bar{\nu}_{a}-E_{Jb}^{2}/\bar{\nu}_{b}){Im}(\mathit{N}%
^{-1}\exp i(\Omega _{V}t-K_{H}x))  \label{Phi}
\end{equation}
where $\mathit{N}=\Omega _{0}^{2}-\Omega _{V}(\Omega _{V}+i\gamma
)+(K_{H}v_{cm})^{2}.$ The phase perturbation $\varphi $ leads to a change of
the dc Josephson current

\begin{equation}
\langle j_{J}\rangle =\langle j_{Ja}\sin (\Omega _{V}t-K_{H}x+\delta (\bar{%
\chi}-\chi _{S})+\varphi /2)+j_{Jb}\sin (\Omega _{V}t-K_{H}x+\delta (\bar{%
\chi}-\chi _{S})-\varphi /2)\rangle  \label{JosCurrent}
\end{equation}
$\delta \bar{\chi}$ is the perturbation of the phase difference between the
total phase $\bar{\chi}$ of the two-band superconductor and the phase $\chi
_{S}$ of the single band superconductor. The angle brackets mean the
averaging in space and time. Then we expand the currents $j_{Ja}$ with
respect to $\delta (\bar{\chi}-\chi _{S})$ and $\varphi /2$ and take into
account Eq.(\ref{Phi}). The correction to the current $j_{J}$ due to the
perturbation $\delta (\bar{\chi}-\chi _{S})$ is omitted because it is not
related to the CMs (this correction is real Fiske steps \cite%
{Abrikosov,Kulik,Barone}) and corresponds to much larger voltages (see
Appendix). The correction to the current $j_{J}$ due to $\varphi $ is

\begin{equation}
\langle \delta j_{J}\rangle =-(j_{Ja}-j_{Jb})(E_{Ja}^{2}/\bar{\nu}%
_{a}-E_{Jb}^{2}/\bar{\nu}_{b})\frac{\gamma \Omega _{V}}{|\mathit{N}|^{2}}
\label{DeltaCurrent}
\end{equation}
where $|\mathit{N}|^{2}=(\Omega _{0}^{2}-\Omega
_{V}^{2}+(K_{H}v_{cm})^{2})^{2}+(\gamma \Omega _{V})^{2}$.

We see that if the applied voltage satisfies the condition $\Omega
_{V}^{2}=\Omega _{0}^{2}+(K_{H}v_{cm})^{2},$ a spike arises on the I-V
characteristics the position of which is shifted with varying magnetic field
$H$. These spikes for different $H$ are shown in Fig.3.

\begin{figure}[tbp]
\par
\begin{center}
\includegraphics[width=0.7\textwidth]{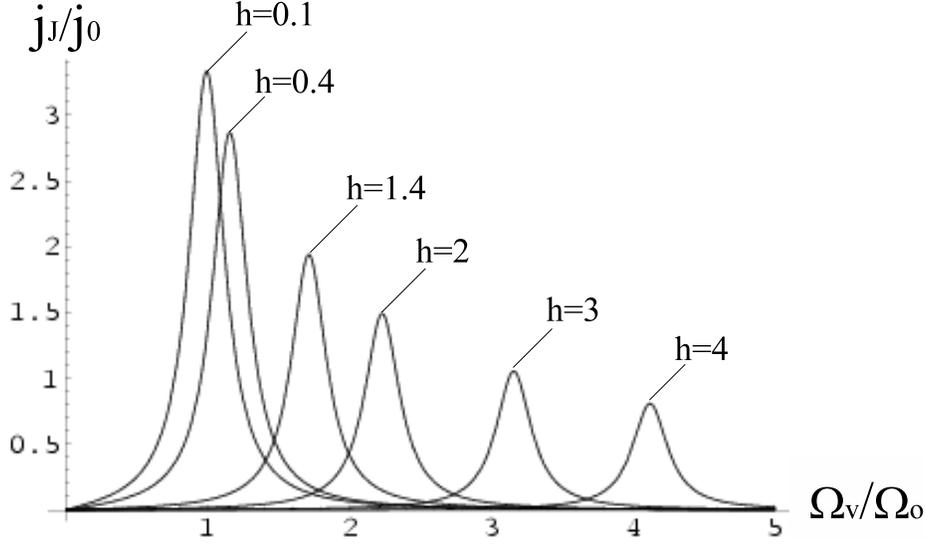}
\end{center}
\caption{Corrections to the I-V characteristics of a Josephson tunnel
junction due to CMs in the two-band superconductor. The Josephson junction
consists of a one-band and two-band superconductors. At voltages $V$
satisying the equation $\Omega _{V}^{2}=\Omega _{0}^{2}+(K_{H}v_{cm})^{2}$ a
spike arises on the I-V characteristics. The position of the spike depends
on an applied magnetic field $H$. The current is normalized to the value: $%
j_{0}=\protect\gamma(j_{Ja}-j_{Jb})(E_{Ja}^{2}/\bar{\protect\nu}%
_{a}-E_{Jb}^{2}/\bar{\protect\nu}_{b})$.}
\end{figure}

\section{Conclusions}

\bigskip

In summary, using a simple model we have studied the CMs in
two-band superconductors and shown that weakly damped CMs exist in
these superconductors at all temperatures below $T_{c\text{ }}$in
the frequency range: $\gamma <\Omega <\Delta _{a,b}$, where
$\gamma $ is an effective charge imbalance relaxation rate. At low
temperatures these modes have a spectrum similar to the spectrum
of the Josephson ``plasma modes''\ in tunnel Josephson junctions.
However, the velocity of the CM in two-band superconductors \ by
the order of magnitude is \ $\sim \sqrt{D\Delta }$, i.e., much
smaller than the velocity of plasma modes in the Josephson
junction.

At high temperatures ($T>>\Delta _{a,b}$), the CM spectrum
consists of two branches. One branch is analogous to the ``plasma
modes''\ in a Josephson junction but propagates much slower than
the Swihart waves. Another branch is similar to the
Carlson-Goldman mode in one-band ordinary superconductors, i. e.,
it has a sound-like spectrum. In the case of intermediate
temperatures ($\Delta _{b}<T<\Delta _{a}$) the gapless
(sound-like) mode has a low damping if strict conditions are
satisfied.

The spectrum of the CMs considered above can be determined experimentally by
measuring the I-V characteristics in a Josephson junction. Such a method has
been applied by Carlson and Goldman [12] for measuring the CM spectrum in a
conventional superconductor. They observed a small peak on the I-V curve the
position of which depends on a weak applied magnetic field. In this case the
wave vector $k$ is proportional to $H$ and the frequency of phase
oscillations $\Omega $ is related to the applied voltage: $\Omega =(2e/\hbar
)V_{B}$. The measured temperature dependence of the spectrum of the CMs in
two-band superconductors would allow one to elucidate the nature of
superconductivity in such two-band superconductors as $MgB_{2}$.

Note that an evidence in favour of the existence of a Leggett-type
collective mode in $MgB_{2}$ was obtained in a recent work, where
the point-contact spectrosopy was used \cite{Maksimov}.

\bigskip

\section{Acknowledgements}

\bigskip

We are grateful to A. Varlamov for attracting our attantion to this problem
and to A.\ Koshelev for useful suggestions. We would like to thank SFB 491
for financial support. A.F.V. also thanks DFG for financial support within
Mercator-Gastprofessoren.

\section{Appendix}

\bigskip .

Here we derive the relation between the phase difference $\theta $ and the
applied voltage $V$ as well as the magnetic field $H$ in the main
approximation, i.e., in the absence of the Josephson coupling. In addition
we present the derivation of Eq.(44). For simplicity we restrict ourselves
with the case of low temperatures ($T<<\Delta _{a,b}$). The results for
higher temperatures are qualitatively the same.

In the absence of the Josephson coupling the right-hand side of Eqs.(40) is
equal to zero. Summing up these equations, we obtain the equation of the
charge conservation

\begin{equation}
e\bar{\nu}_{a,b}\partial \mu _{a,b}/\partial t=(\bar{\nu}D/\sigma
)_{a,b}\nabla \mathbf{j}_{a,b}  \label{A1}
\end{equation}
where $j_{a,b}=\alpha _{a,b}(\partial _{x}\chi _{a,b}-2\pi A_{x}/\Phi
_{0})/2;$ $\Phi _{0}=hc/2e$ is the magnetic flux quantum. The coefficients $%
\alpha _{a,b}$ are related to London penetration depth $\lambda _{L}$: $%
\alpha _{a,b}=c\Phi _{0}/(4\pi ^{2}\lambda _{La,b}^{2}).$ At low
temperatures these coefficients are equal to $\alpha _{a,b}=(\sigma \Delta
)_{a,b}/e$. Summing up these equations, we obtain the equations of the
charge conservation

\begin{equation}
e\partial (\bar{\nu}_{a}\mu _{a}+\bar{\nu}_{b}\mu _{b})/\partial t=\nabla
\mathbf{[\alpha }_{a}\mathbf{Q}_{a}+\mathbf{\alpha }_{b}\mathbf{Q}_{b}%
\mathbf{]}  \label{A2}
\end{equation}
where $Q_{a,b}=(\partial _{x}\chi _{a,b}-2\pi A_{x}/\Phi _{0})/2$ is the
condensate momentum. Differentiating this equation on time and taking into
account Eq.(14) and the Poisson equation
\begin{equation}
\nabla E=4\pi \rho =4\pi \upsilon e(\bar{\nu}_{a}\mu _{a}+\bar{\nu}_{b}\mu
_{b})  \label{A3}
\end{equation}
we obtain the equation

\begin{equation}
v^{2}\partial ^{2}(\bar{\nu}_{a}\mu _{a}+\bar{\nu}_{b}\mu _{b})/\partial
t^{2}=-k_{TF}^{2}(\bar{\nu}_{a}\mu _{a}+\bar{\nu}_{b}\mu _{b})+\nabla ^{2}[%
\bar{\delta}_{a}\mu _{a}+\bar{\delta}_{b}\mu _{b}\mathbf{]}  \label{A4}
\end{equation}
where $v^{2}=(\pi /2)[(D\bar{\nu}\Delta )_{a}+(D\bar{\nu}\Delta )_{b}],\nu
=\nu _{a}+\nu _{b}$, $\bar{\delta}_{a,b}=(\sigma \Delta )_{a,b}/((\sigma
\Delta )_{a}+(\sigma \Delta )_{b}),$ $k_{TF}^{2}=4\pi e^{2}\nu $. The
Thomas-Fermi screening length $k_{TF}^{-1}$ is of the order of interatomic
spacing, i.e., is much shorter than charactetistic lengths of the problem: $%
k_{TF}^{{}}>>v/\Delta >v/\Omega _{V}\sim K_{H}.$ This means that the first
term on the right-hand side in Eq.(\ref{A4}) is much larger than other
terms. Therefore we obtain

\begin{equation}
\bar{\nu}_{a}\mu _{a}+\bar{\nu}_{b}\mu _{b}=0  \label{A5}
\end{equation}
For the single band superconductor we have

\begin{equation}
\mu _{S}=0  \label{A6}
\end{equation}

Writing the potentials $\mu _{a,b}$ in the form $\mu _{a,b}=\partial (\bar{%
\chi}\pm \varphi /2)/\partial t+eV,$ we obtain from Eqs.(\ref{A5}-\ref{A6})

\begin{equation}
\partial \theta /\partial t=2eV-\frac{1}{2}(\bar{\nu}_{a}-\bar{\nu}%
_{b})\partial \varphi /\partial t  \label{A7}
\end{equation}
where $\theta =\bar{\chi}-\chi _{S},V_{B}=V-V_{S}$ is the applied voltage.
Eq.(\ref{A7}) generalizes \ the Josephson relation to the case of a junction
with a two-band superconductor.

If in the ground state there are no $\varphi $-solitons ($\varphi =0)$ and
the external magnetic field $H_{e}$ is applied in the $y$-direction, one
obtains from Eq.(\ref{A7})

\begin{equation}
\theta (x,t)=\Omega _{V}t-K_{H}x  \label{A8}
\end{equation}%
where $\Omega _{V}=2eV_{B}/\hbar $ and the wave vector $K_{H}$ is determined
by the equation \cite{Kulik}

\begin{equation}
K_{H}=(2\pi H_{e}/\Phi _{0})[\lambda \tanh (d/\lambda )+\lambda _{S}\tanh
(d_{S}/\lambda _{S})]  \label{A9}
\end{equation}
where $d_{S}$ is the thickness of the single band superconductor, $\lambda
=\lambda _{La}+\lambda _{Lb}$.

We turn now to finding solutions for Eqs.(\ref{EqPhi1}-\ref{EqPhi2}). We
assume that the phase perturbations are small ($\chi _{a,b},\varphi <<1$)
and seek for a solution in the form

\begin{equation}
\varphi (x,t)={Im}\varphi _{\Omega }\exp (\Omega _{V}t-K_{H}x)  \label{A10}
\end{equation}

Summing up Eqs.(\ref{EqPhi1}-\ref{EqPhi2}), we obtain

\begin{equation}
-\partial (\bar{\nu}_{a}\mu _{a}+\bar{\nu}_{b}\mu _{b})/\partial t+\partial
\lbrack \alpha _{a}Q_{a}+\alpha _{b}Q_{b}]/\partial x=E_{Ja}^{2}+E_{Jb}^{2}
\label{A11}
\end{equation}

The first term on the left is small due to the quasineutrality condition.
Therefore we obtain from Eq.(\ref{A11})

\begin{equation}
Q_{b}=-\frac{(\sigma \Delta )_{a}}{(\sigma \Delta )_{b}}%
Q_{a}+(E_{Ja}^{2}+E_{Jb}^{2})/(iK_{H}v_{b}^{2}\bar{\nu}_{b})  \label{A12}
\end{equation}

From the definition of $Q_{a,b}$ we have

\begin{equation}
Q_{a}-Q_{b}=\frac{1}{2}\partial \varphi /\partial x  \label{A13}
\end{equation}

Finally we divide Eqs.(\ref{EqPhi1}-\ref{EqPhi2}) by $\bar{\nu}_{a,b}$ and
subtract from each other. We get

\begin{equation}
2\epsilon _{0}^{2}(\bar{\nu}_{a}^{-1}+\bar{\nu}_{b}^{-1})\varphi -\frac{1}{2}%
\partial ^{2}\varphi /\partial t^{2}+2v_{a}^{2}\bar{\nu}_{a}\partial
Q_{a}/\partial x=2E_{Ja}^{2}  \label{A14}
\end{equation}

Using Eq.(\ref{A13}), we substitude $Q_{a}$ into Eq.(\ref{A14}) and come to
Eq.(\ref{Phi}) for $\varphi $.

\end{document}